\documentclass[aps,prb,twocolumn,superscriptaddress,amssymb,citeautoscript,showkeys,floatfix]{revtex4-2}
\pdfoutput=1
\usepackage{dsfont}
\usepackage{graphicx} 
\usepackage{dcolumn}% Align table columns on decimal point 
\usepackage{bm}% bold math
\renewcommand{\vec}[1]{\bm{#1}} 
\usepackage{times} 
\usepackage{array} 

\begin{document}

\title{Current-controlled chirality dynamics in a mesoscopic magnetic domain wall}

\author {Oleksiy Kolezhuk} 
\affiliation{Institute of Physics, Johannes Gutenberg-University Mainz, D-55128 Mainz, Germany} 
\affiliation{Institute of High Technologies, Taras Shevchenko National University of Kyiv, 03022 Kyiv, Ukraine} 
\affiliation{Institute of Magnetism, National Academy of Sciences and Ministry of Education and Science, 03142 Kyiv, Ukraine}

\author {Roman Teslia} \author{Ihor Buryak} 
\affiliation{Institute of High Technologies, Taras Shevchenko National University of Kyiv, 03022 Kyiv, Ukraine}

\author {Olena Gomonay} 
\affiliation{Institute of Physics, Johannes Gutenberg-University Mainz, D-55128 Mainz, Germany}

\begin{abstract} 
Chirality as internal degree of freedom of a mesoscopic domain wall inside a quasi-one-dimensional fixture can be controlled by spin-polarized current for ferro- as well as antiferromagnetic domain walls. 
We show that the current
density required for the chirality manipulation can be significantly reduced in the low-temperature regime where the chirality dynamics exhibits quantum effects. In this quantum regime, weak currents can excite Bloch oscillations of the domain wall angular rotation velocity, with the oscillation frequency proportional to the current, modulated by a much higher magnon-range frequency. 
In addition to that, the Wannier-Stark localization effects enable controlled switching between different chiral states, suppressing inertial effects characteristic for the classical regime.
We also show that for recently discovered novel class of magnetic
materials - altermagnets - chirality switching can be driven by the usual charge current (not spin-polarized). 
\end{abstract} 

\maketitle

\section{Introduction} 
\label{sec:intro}

The last decade has witnessed great theoretical and experimental progress in spintronics,  significantly extending our capabilities of generation,
control, manipulation, and detection of spin textures by current (see, e.g., \cite{Manchon2018,Jungwirth+16rev}). 
This progress is in particular driven by rich possibilities offered by effects of spin-orbit coupling \cite{Nagaosa2010,Manchon2018} and fast magnetization dynamics in antiferromagnets \cite{Gomonay2014b,Gomonay2016,Gomonay2017,Jungwirth+16rev,Baltz+18rev,Chen+22rev}, with realization of all-electric fast switching \cite{Zelezny2014,Wadley2016,Bodnar2018} bearing promise for non-volatile storage operating in terahertz domain. In altermagnets, recently discovered class of magnetic materials that combine time-reversal symmetry breaking with collinear antiferromagnetic spin order,  large non-relativistic spin-momentum coupling opens avenues to efficient spin current injection, robust giant magnetoresistance and other nontrivial effects \cite{Smejkal+22rev,Smejkal2020,Feng2022,Smejkal2022-prx}.

As the physics of magnetic and spintronic devices advances towards ever
smaller scales, spin textures, usually considered as classical fields, can acquire quantum behavior. Those quantum aspects received a good deal of attention in the past, in the context of the quantum tunneling of spin in magnetic nanoparticles \cite{ChudnovskyTejada-book05}, molecular clusters \cite{Gatteschi+book06}, and topologically nontrivial magnetic textures \cite{Ivanov05rev}.  Understanding the quantum dynamics of spin textures interacting with spin or charge currents would provide a foundation for merging and hybridizing quantum and classical computing technologies.

In this paper, we focus on the study of a current-driven dynamics of a magnetic domain wall localized inside a nanosized quasi-one-dimensional structure (wire, stripe, constriction, etc), which has an internal degree of freedom, chirality, characterizing the way of
rotation of magnetization inside the wall. Classically,  two states with opposite chirality are equivalent in energy, while quantum-mechanically, there is a finite tunneling amplitude mixing the two states and lifting the degeneracy. In the absence of current, such a macroscopic quantum tunneling of magnetization inside a domain wall has been studied in a number of works \cite{IK94-e,IK95tunn-e,BraunLoss96,TakagiTatara96,IKK98}.
 
On the other hand, the interaction of current with such a localized domain wall has been studied only at the classical level.
It has been shown  that for domain walls in ferromagnets (FM) \cite{Berger1986,%%Thomas2006,HeZhang2007,
OnoNakatani2008,Bisig+2009,Martinez+2011} and antiferromagnets (AF) \cite{ChengNiu14,Ovcharov+22}, spin-polarized current directly couples to the chirality degree of freedom, and the corresponding torque can drive domain wall rotation. 

Here we revisit the problem of current-driven chirality dynamics at the quantum level. We demonstrate that  in AF and in altermagnets (AM), in the low-temperature regime when the chirality dynamics acquires quantum features, the current density required for efficient manipulation of chirality can be significantly reduced in comparison with the classical regime realized at high temperatures. 
In particular, we show that weak currents can trigger quantum Bloch oscillations modulated by a much higher magnon-range frequency, a regime which is very different from the classical rotation. Moreover, Zener breakdown can be driven by currents below the classical threshold, and the Wannier-Stark localization can be exploited to switch chirality without inertial effects, unlike the classical regime.
In addition, we also show that in AM, in the contrast to other magnets, chirality can be manipulated by the unpolarized current. 
 
 The rest of the paper is structured as follows: in Sect.~\ref{sec:model}, we introduce the setup and provide a unified derivation of effective Lagrangians describing interaction of current with chirality for domain walls in FM, AF, and AM; in Sect.~\ref{sec:regimes} we focus on the peculiarities of the chirality dynamics in the quantum regime, and discuss the feasibility of their experimental observation; finally, Sect.~\ref{sec:summary} provides a brief summary.

\section{Model: pinned domain wall in a nanowire} 
\label{sec:model}

\begin{figure}[tb] 
\includegraphics[width=0.38\textwidth]{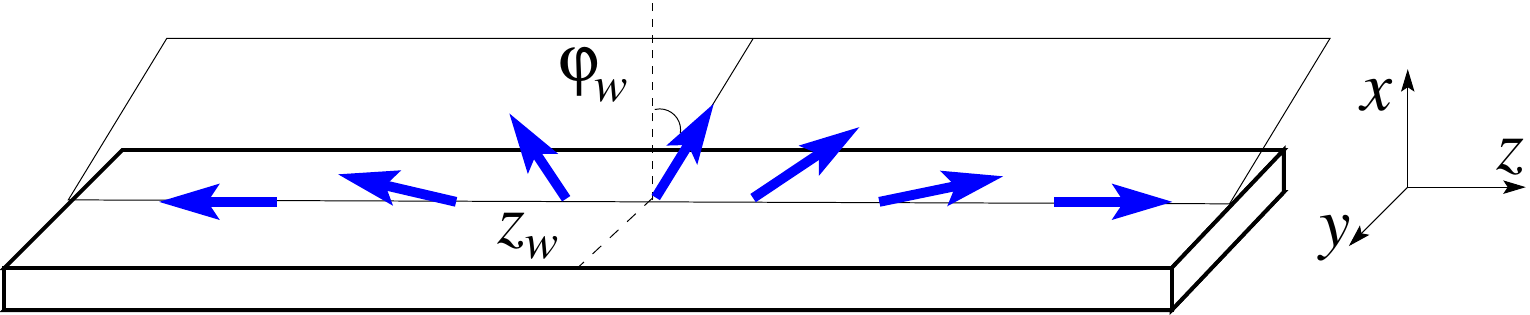} 
\caption{\label{fig:wire} (Color online). Schematic picture of the mesoscopic domain wall in a quasi-one-dimensional fixture. Arrows show the direction of the magnetic order parameter (magnetization or the N{\'e}el vector). } 
\end{figure}

Consider a domain wall (DW) inside a magnetic nanowire as shown in Fig.~\ref{fig:wire}. 
We assume the DW is pinned by some potential of the simplest quadratic form $U_{\text{pin}}=Gz_w^2/2$, where $z_w$ is the position of the DW center.  We will study ferromagnetic as well as antiferromagnetic DWs, denoting the respective order parameter  by the unit vector field $\vec{n}$. It is convenient to use the standard spherical angles parametrization,
$n_x+in_y=\sin\theta e^{i\varphi}$, $n_z=\cos\theta$. The cross-section area $C_\perp$ of the DW contains $N_\perp=C_\perp a/V_0$ magnetic atoms, where $a$ is the magnetic lattice constant along the direction of the wire, and $V_0$ is the volume of magnetic unit cell.

There is exchange coupling $J$ and a biaxial magnetic anisotropy, with the anisotropy energy per unit cell 
\begin{equation} 
\label{anis}
w_a=S^2(K_1 n_x^2 +K_2 n_y^2),
\end{equation} 
where $S$ denotes the underlying spin of magnetic atoms, and the anisotropy constants $K_1>K_2>0$, so
$z$ is the easy axis and $x$ the hard axis. 
In what follows, we will assume that  the anisotropy is close to uniaxial, i.e., the rhombicity parameter
\begin{equation}
\label{rho}
\rho=(K_1-K_2)/K_2 \ll 1.
\end{equation}
Under this latter assumption, one can approximately consider that along the DW spins rotate in a plane with a constant angle $\varphi$, for both the FM and AF domain walls \cite{IK94-e,BraunLoss96,TakagiTatara96}, and describe the DW solution with the help of two collective coordinates, the DW position $z_{w}$ and its angle $\phi_{w}$:
\begin{equation}
\label{ansatz-col}
\cos\theta=\tanh\big((z-z_w)/\delta\big),\quad \varphi=\phi_w,
\end{equation}
where $\delta=a\sqrt{\frac{J}{2K_2}}$ is the DW thickness. The static DW energy is given by
\begin{equation}
\label{DW-en}
E_w=N_\perp E_0 \big( 1+(\rho/2)\cos^2\phi_w \big),\quad E_0=2S^2\sqrt{2JK_2}.
\end{equation}
Two lowest  DW states with $\phi_w=\pm\pi/2$ differ by the sense of rotation of $\vec{n}$ inside the $(yz)$ easy plane. 

In what follows in this section, we discuss the effective Lagrangians describing interaction of current with chirality for pinned domain walls in FM, AF, and AM. Although the results for FM and AF are well-documented in the literature, we revisit them here for the benefit of the reader, to provide a unified context for the respective derivation for AM; technical details are relegated to the Appendix.

\subsection{Domain wall in a ferromagnet}
\label{subsec:FM}

In the case of a FM, $\vec{n}$ is the unit vector of magnetization, and the system is described by the Lagrangian (see, e.g., \cite{assa})
\begin{eqnarray}
\label{Lagr-FM} 
\mathcal{L}&=&-(N_\perp\hbar S/a)\int dz \big(\vec{A}\cdot\partial_t\vec{n}\big) -W[\vec{n}] -U_{\text{pin}}\\ 
W[\vec{n}]&=&
\frac{N_\perp S^2}{a}\int dz \Big\{ \frac{Ja^2}{2}(\partial_z \vec{n})^2 + K_1 n_x^2 +K_2 n_y^2\Big\}.\nonumber 
\end{eqnarray} 
Here $\vec{A}$ is the vector potential of the Dirac monopole,
\begin{equation} 
\label{Adirac} 
\vec{A}(\vec{n}_0,\vec{n})={(\vec{n}_0\times\vec{n})}/\big({1+(\vec{n}_0\cdot\vec{n})}\big). 
\end{equation} 
The direction of the ``Dirac string'' given by the unit vector $\vec{n}_0$ can be chosen arbitrarily, and equations of motion
as well as other physical quantities do not depend on the choice of $\vec{n}_0$. In what follows, we choose $\vec{n}_0=\hat{\vec{z}}$ along the easy axis.

%%%%%%%%%%%%%%%%%%%%%%%%%%%%%%%%%%%% 
%%%%%%%%%%%%%%%%%%%%%%%%%%%%%%%%%%%%

Electrons interacting with an inhomogeneous magnetization texture experience a fictitious ``gauge field'' caused by the texture gradients which couples to spin current \cite{ShraimanSiggia88,Bazaliy98,KohnoShibata07}. 
Performing a unitary transformation rotating the spin quantization axis of conduction electrons from its initial value (assumed to be $\vec{n}_0$)  to the local magnetization direction $\vec{n}$ leads, 
to the first order in the magnetization space-time gradients,  to the following contribution to the Lagrangian (\ref{Lagr-FM}):
\begin{equation}
\label{scurr-cpl1}
\mathcal{L}\mapsto \mathcal{L}-I^{(s)}P_z- \mathcal{T}^{(s)}P_0 ,\quad P_\mu=\int dz\,\vec{A}(\hat{\vec{z}},\vec{n})\cdot \partial_\mu\vec{n},
\end{equation}
where $I^{(s)}$ is the spin current magnitude, and $\mathcal{T}^{(s)}$ is the net spin density of the electrons. 
For spin-polarized current, both $I^{(s)}$ and  $\mathcal{T}^{(s)}$ are proportional to the charge current $I^{(c)}$:
\begin{equation}
\label{Is-Ts}
I^{(s)}={\hbar \mathcal{P} I^{(c)}}/{2e},\quad \mathcal{T}^{(s)}= {I^{(s)}}/{v_s},
\end{equation}
where $\cal{P}$ is the degree of spin polarization of the electric current and $v_s$ is the effective electron velocity.
The contribution proportional to $P_0$ is merely a weak renormalization of the first term in the Lagrangian (\ref{Lagr-FM}), so in a FM it can be neglected. The integral $P_z$ can be obtained (see the Appendix) as 
\begin{equation}
\label{Pw}
P_z=2\phi_w.
\end{equation}
 
The effective Lagrangian for a ferromagnetic DW is thus obtained in the following form:
\begin{eqnarray}
\label{Ldw-FM}
\mathcal{L}_{FM}&=&\frac{2N_\perp \hbar S}{a}z_{w}\frac{d\phi_w}{dt}-N_\perp E_0 \big( 1+ (\rho/2)\cos^2\phi_w \big) \nonumber\\
&-&2I^{(s)} \phi_w - {G z_w^2}/{2}.
\end{eqnarray}
The corresponding Hamiltonian is 
\begin{equation}
\label{Hdw-FM}
\mathcal{H}_{FM}=\frac{P_\phi^2}{2M_\phi }+N_\perp E_0 \Big(1+\frac{\rho}{2}\cos^2\phi_w\Big) 
+2I^{(s)} \phi_w ,
\end{equation}
where $P_\phi=\frac{2N_\perp \hbar S}{a} z_w$ is the canonical momentum, and
\begin{equation}
\label{M-phi}
M_\phi={(2N_\perp \hbar S)^2}/{Ga^2}
\end{equation}
plays the role of the effective moment of inertia of the DW. In the absence of current,  small oscillations around the equilibrium (given by $z_w=0$, $\phi_w=\pm\pi/2$) have the frequency 
\begin{equation}
\label{Omega0}
\Omega_0=(N_\perp E_0\rho/M_\phi )^{1/2}. 
\end{equation}

%%%%%%%%%%%%%%%%%%%%%%%%%%%%%%%%%

\subsection{Antiferromagnetic domain wall}
\label{subsec:AF}

For an AF, the dynamics of the unit N\'eel vector $\vec{n}$ is described by the Lagrangian of the nonlinear sigma-model
\begin{equation}
\label{Lagr-AF} 
\mathcal{L} =  \frac{1}{2c^2}N_\perp JS^2a\int dz \,(\partial_t\vec n)^2 -W[\vec{n}]-U_{\text{pin}},
\end{equation}
where $c$ is the limiting velocity of spin waves. 
The spin-transfer torque caused by the interaction of conduction electrons with the AF spin texture can be derived in a way very similar to the FM case \cite{ShraimanSiggia88, SwavingDuine11,ChengNiu12,ChengNiu14}, and the corresponding contribution to the Lagrangian has the same form (\ref{scurr-cpl1}). It should be remarked that non-adiabatic contributions to spin transfer torque in an antiferromagnet have been shown \cite{Fujimoto21} to lead to the coupling of same form, albeit multiplied by a non-universal factor.
The effective Langangian for the antiferromagnetic DW is
\begin{eqnarray}
\label{Ldw-AF}
L_{AF}&=& \frac{N_\perp E_0}{2c^2} \left\{ \Big(\frac{dz_w}{dt}\Big)^2 +\delta^2 \Big(\frac{d\phi_w}{dt}\Big)^2\right\} \nonumber\\
&-&  N_\perp E_0 \Big( 1 + (\rho/2) \cos^2\phi_w \Big) - G z_w^2/2\nonumber\\
&-&2I^{(s)} \phi_w + 2\mathcal{T}^{(s)}z_w\frac{d\phi_w}{dt}.
\end{eqnarray} 

Note that in the AF case the contribution proportional to $\mathcal{T}^{(s)}$ is the only source of coupling between the DW position and angle \cite{ChengNiu14} and thus cannot be written off as a small correction to an existing term. However, for strong pinning, when the characteristic frequency 
\begin{equation}
\label{freq-pin-af}
\Omega_{\text{pin}}=(Gc^2/N_\perp E_0)^{1/2}
\end{equation}
of $z_w$ oscillations is much larger than the  frequency
\begin{equation}
\label{omega0}
\omega_0=(E_0/\hbar S )\sqrt{\rho}.
 \end{equation}
of $\phi_w$ oscillations  around equilibrium points $\phi_w=\pm\pi/2$ , the DW coordinate can be treated as ``slave'' and integrated out, $z_w\simeq (2\mathcal{T}^{(s)}/G)(d\phi_w/dt)$, which leads just to a weak renormalization of the limiting velocity $c$.

Moreover, since the last term in (\ref{Ldw-AF}) can be recast in the equivalent form $-2\mathcal{T}^{(s)}\phi_w\frac{dz_w}{dt}$, it is easy to see that it contains the additional  parameter $v_s^{-1}(dz_w/dt)$ compared to the last but one term in the same expression. If the DW velocity is much smaller than the effective electron velocity $v_s$ (which is estimated to be in the range of $1$ to $100$~km/s \cite{ChengNiu14}), this coupling can be neglected.  
Thus, in contrast to the FM case, for small rhombicity $\rho\ll1$ the translational motion of the AF domain wall (dynamics of $z_w$) is only weakly coupled to its chirality (dynamics of $\phi_w$). 

The Hamiltonian describing the $\phi_w$ dynamics in an AF DW can thus be written as 
\begin{equation}
\label{Hdw-AF}
\mathcal{H}_{AF}=\frac{p_\phi^2}{2m_\phi}+N_\perp E_0 \Big( 1 + \frac{1}{2}\rho \cos^2\phi_w \Big)+2I^{(s)} \phi_w,
\end{equation}
where $p_\phi=m_\phi (d\phi_w/dt)$ is the canonocal momentum, andy 
\begin{equation}
\label{m-phi}
m_\phi=N_\perp E_0(\delta/c)^2
\end{equation}
is the effective DW moment of inertia. 

%%%%%%%%%%%%%%%%%%%%%%%%%%%%%%%%%%%% 
\begin{figure}[tb] 
\includegraphics[width=0.16\textwidth]{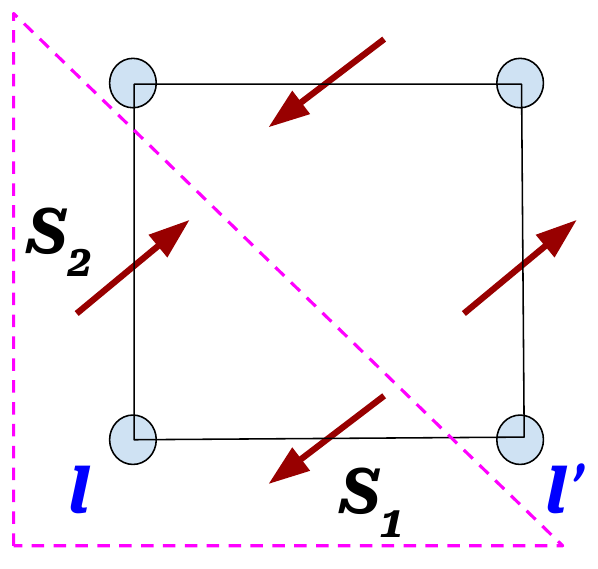} 
\caption{\label{fig:sqlatam}  (Color online).
Illustration of the toy model of altermagnet described by Eq.\ (\ref{tight-am}). 
Arrows denote localized spins, and the dashed line indicates the magnetic unit cell. } 
\end{figure} 
%%%%%%%%%%%%%%%%%%%%%%%%%%%%%%%%%%%%

\subsection{Domain wall in an altermagnet}
\label{subsec:AM}

The low-energy spin dynamics in an altermagnet in the leading approximation is described by the same nonlinear sigma-model as in the antiferromagnetic case (we neglect slight corrections caused by the back action of the electron subsystem on the magnetization).
However, one can show that, in contrast to the FM or AF, in an altermagnet the gradient of the AF order parameter couples not only to spin current, but to the charge current as well. This can be easily seen in the simplest tight-binding model of an altermagnet \cite{Smejkal+22rev}, described by the Hamiltonian
\begin{equation}
\label{tight-am}
\hat{H}=\sum_{\langle ll' \rangle}\Big\{ -t \psi_l^\dagger \psi_{l'} - t' \psi_l^\dagger (\vec{\sigma}\cdot\vec{S}_{ll'}) \psi_{l'} +\text{h.c.}\Big\},
\end{equation}
where $l$ labels sites on a square lattice,  $t$ and $t'$ are the amplitudes of spin-independent and spin-dependent hopping, respectively, and localized spins $\vec{S}_{ll'}$ sit in the middle of the links between nearest neighbor pairs $\langle ll' \rangle$, as shown in Fig.\ \ref{fig:sqlatam}. Although this  toy model is just one of many possible realizations of an altermagnet, it allows us to capture the essential physics of AMs, so we expect the final results to remain qualitatively correct in the general case. 

 Introducing in each unit cell the magnetization $\vec{m}$ and the N\'eel vector $\vec{n}$ in a standard way,  and rotating the electron quantization axes to the local N\'eel vector direction, one can see that the spin-independent  hopping again yields the coupling of texture gradients to the spin current as given by Eq.\ (\ref{scurr-cpl}), while the spin-dependent hopping leads to the following contribution to the Lagrangian density in the continuum:
\begin{equation}
\label{am-coupling}
-\kappa\frac{\hbar}{2e}
  \vec{A}(\hat{\vec{z}},n)\cdot \Big(  I^{(c)}_{x} \partial_x \vec{n} - I^{(c)}_{y}   \partial_y \vec{n} \Big) ,
\end{equation}
where  $\kappa=t'/t$ is the relative strength of the spin-dependent hopping, and  $\vec{I}^{(c)}$ is the charge current,
see (\ref{alt-coupling}) and the corresponding derivation in the Appendix. 
 
Thus, in the DW setup as considered above, with the nanowire oriented along along one of the directions $x$ or $y$, with an electric current of the magnitude $I^{(c)}$   and spin-polarization $\mathcal{P}$ flowing along the nanowire, one obtains 
the Hamiltonian of the form (\ref{Hdw-AF}), with the replacement
\begin{equation}
\label{am-cpl}
I^{(s)} \mapsto \frac{\hbar I^{(c)}}{2e}(\mathcal{P}\pm \kappa).
\end{equation}

%%%%%%%%%%%%%%%%%%%%%%%%%%%%%%%%%
\section{Classical and quantum chirality control} 
\label{sec:regimes}
%%%%%%%%%%%%%%%%%%%%%%%%%%%%%%%%% 

In all cases considered above the DW is described by the effective Hamiltonian of the following form: 
\begin{equation}
\label{hamgen}
\mathcal{H}=\frac{p^2}{2m}+\frac{1}{2}m\omega^2\cos^2\phi_w + b \phi_w,
\end{equation}
where the moment of inertia $m$ and the frequency $\omega$ are given by $m=M_\phi$, $\omega=\Omega_0$ for FM, , see Eqs.\ (\ref{M-phi}), (\ref{Omega0}), and $m=m_\phi$, 
$\omega=\omega_0$ for AF/AM, see Eqs.\ (\ref{m-phi}), (\ref{omega0}). 
The Hamiltonian (\ref{hamgen}) describes a planar rotator in a two-well potential.
The two wells are separated by the barrier of the height 
\begin{equation}
\label{barrier}
U={m\omega^2}/{2}=\rho E_0 N_\perp/2.
\end{equation}
This planar rotator is driven by the torque  $b=\gamma  \hbar I^{(c)}/e$  proportional to the current $I_c$, with the coefficient $\gamma=\mathcal{P}$ for FM and AF, and $\gamma =\mathcal{P}\pm\kappa$ for AM.  
Applying a constant or time-dependent torque, one can excite the DW chiral degree of freedom in various ways. 

The above effective model is considerably simplified, as it omits effects of non-adiabaticity as well as dissipation. However, we think it represents an acceptable starting point for our primary goal of exploring the quantum chirality dynamics.
We first give a  brief overview of the classical dynamics that is well studied in the literature, and then focus on the quantum regime which, as we shall see, exhibits a number of novel features. 

%%%%%%%%%%%%%%%%%%%%%%%%%%%%%%%%%%%% 
\begin{figure}[tb] 
\includegraphics[width=0.45\textwidth]{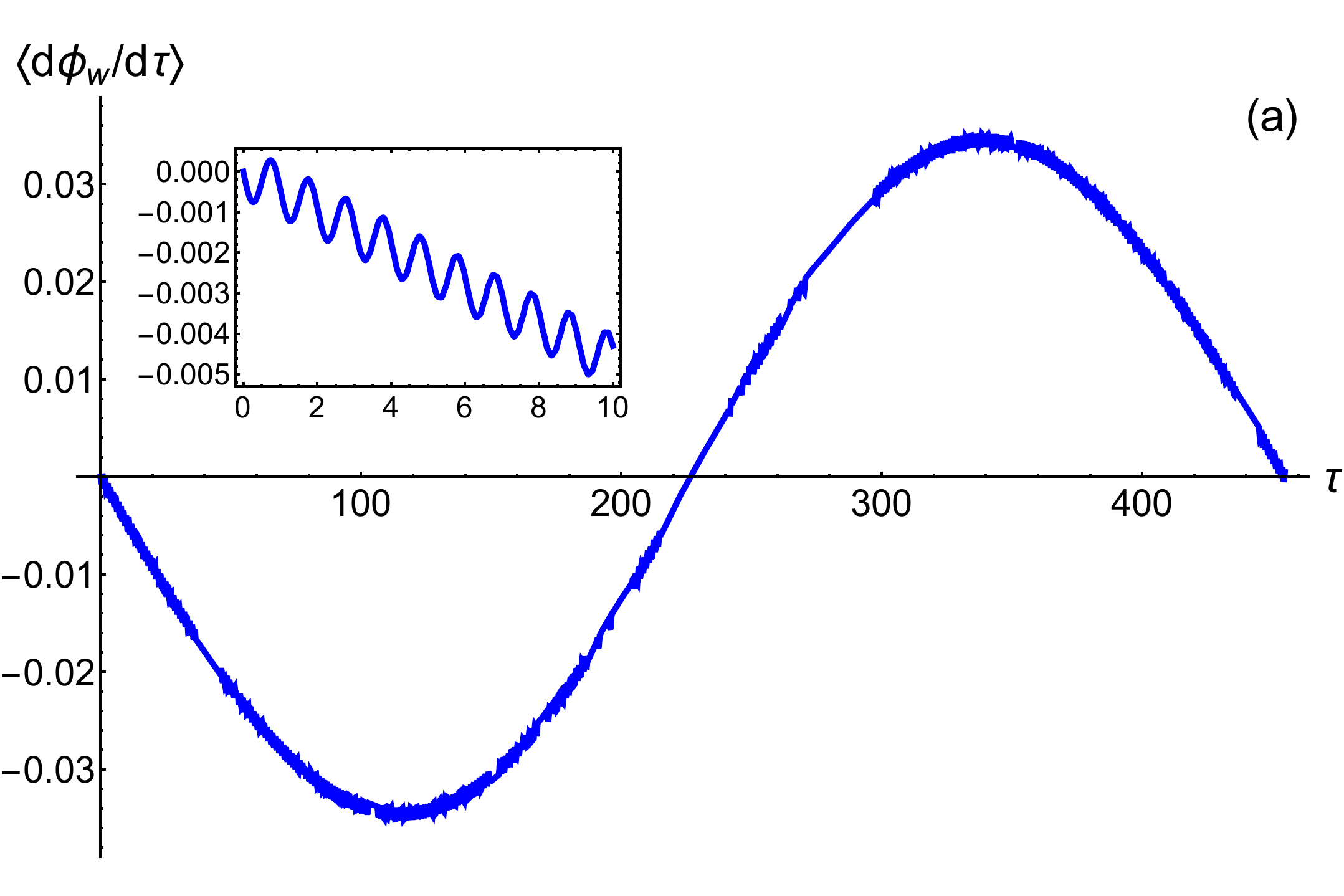}

\includegraphics[width=0.45\textwidth]{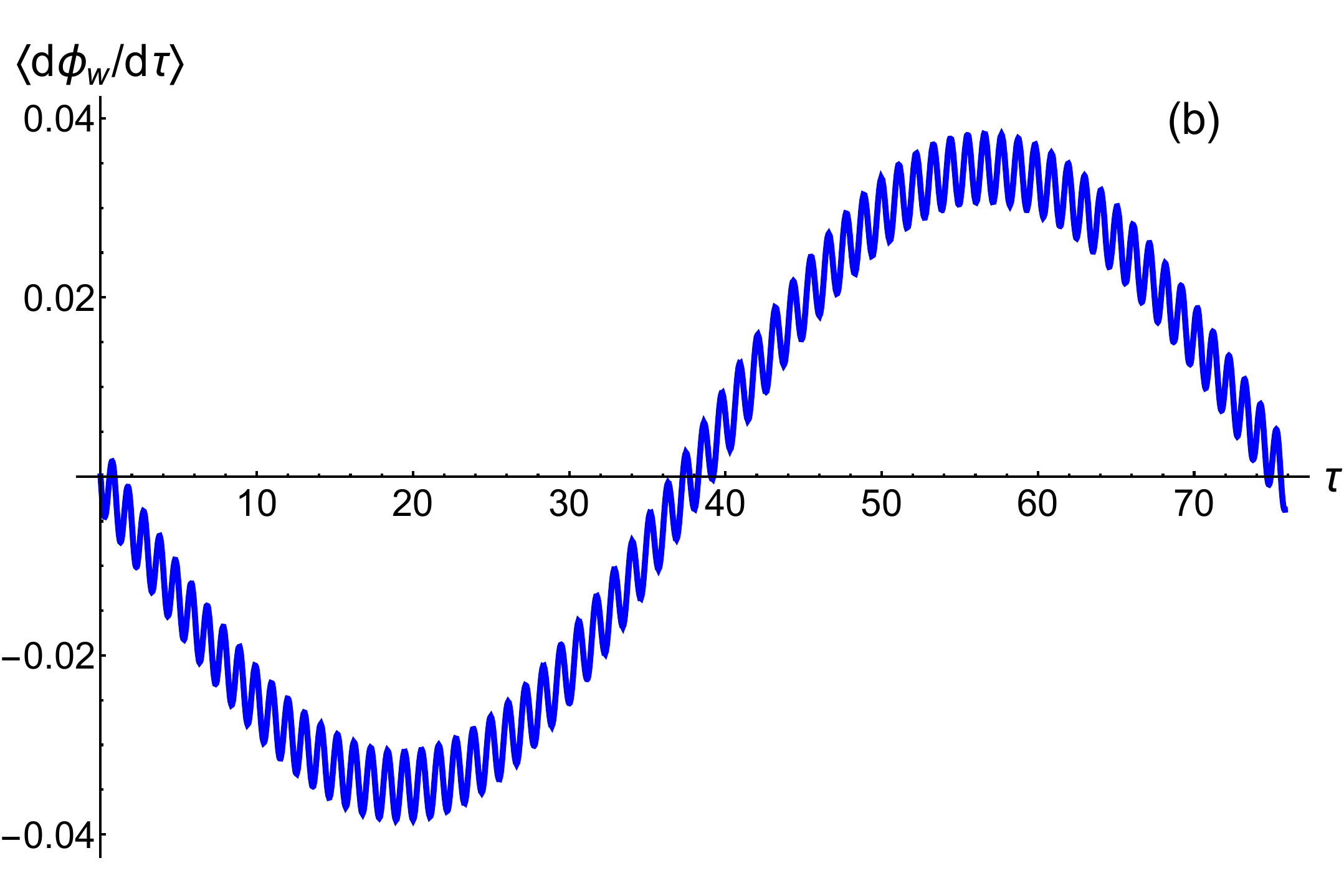}

\caption{\label{fig:bloch1s} 
 (Color online).
The average angular velocity $\langle d\phi_w/d\tau\rangle$  vs the dimensionless time $\tau=\hbar t/2m$
 for $\xi=5$, and bias (a) $b=0.1\Delta$; (b) $b=0.6\Delta$. The ground state for $b=0$ has been used as the initial state at $t=0$. 
 The energies of the first four lowest-lying states at $\xi=5$ in units of $\hbar^2/2m$ are $E_0\approx 0.115$, $E_1\approx 0.159$, $E_2\approx 5.636$, and $E_3\approx 6.343$. 
 Bloch oscillations with the frequency $\pi b\hbar$ are modulated with the secondary frequency  $(E_3-E_0)/\hbar$ due to the excitation of the third excited state.
 } 
\end{figure} 
%%%%%%%%%%%%%%%%%%%%%%%%%%%%%%%%%%%%

%%%%%%%%%%%%%%%%%%%%%%%%%%%%%%%%%%%% 
\begin{figure*}[tb] 
\includegraphics[width=0.45\textwidth]{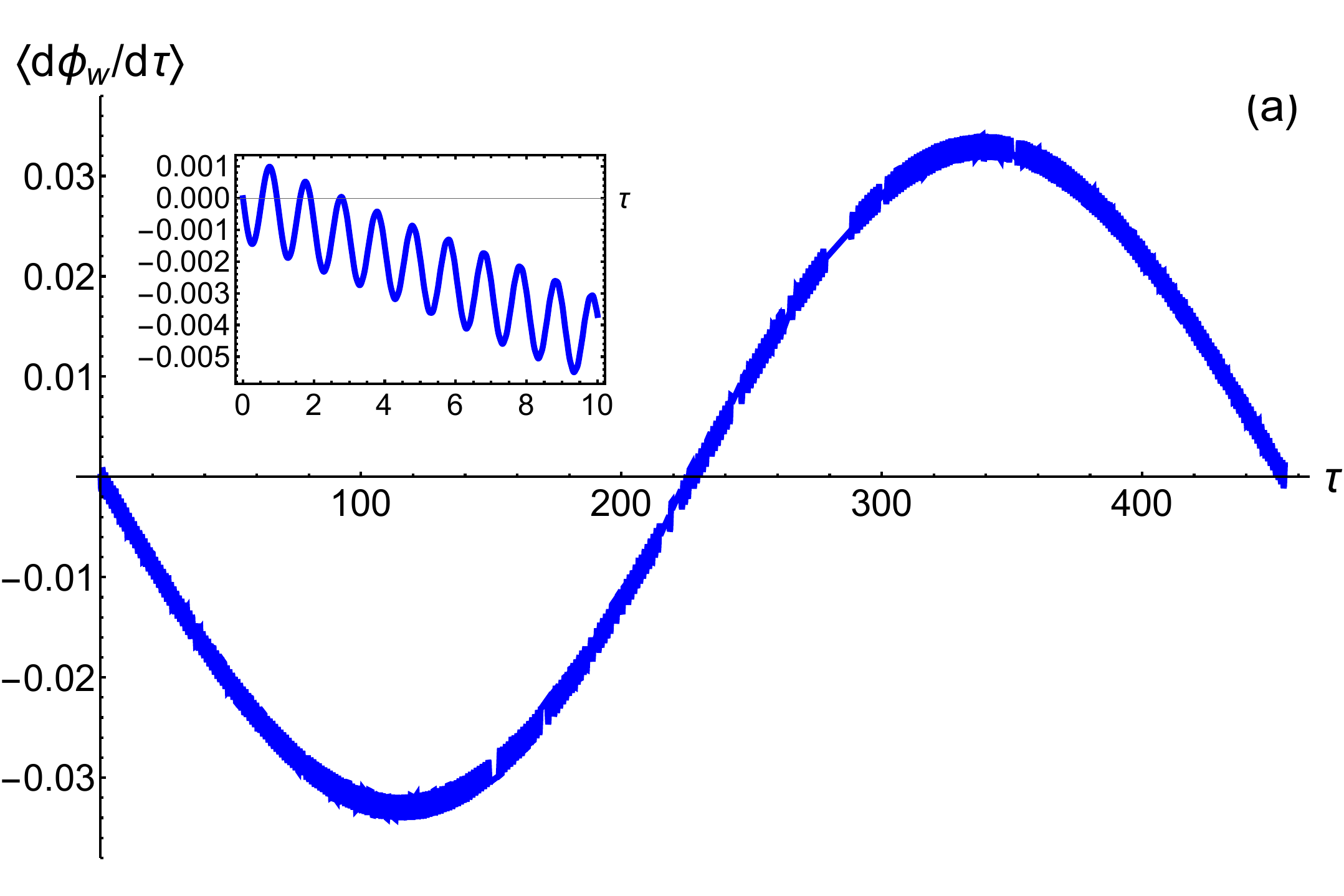}
\hspace*{5mm} \includegraphics[width=0.31\textwidth]{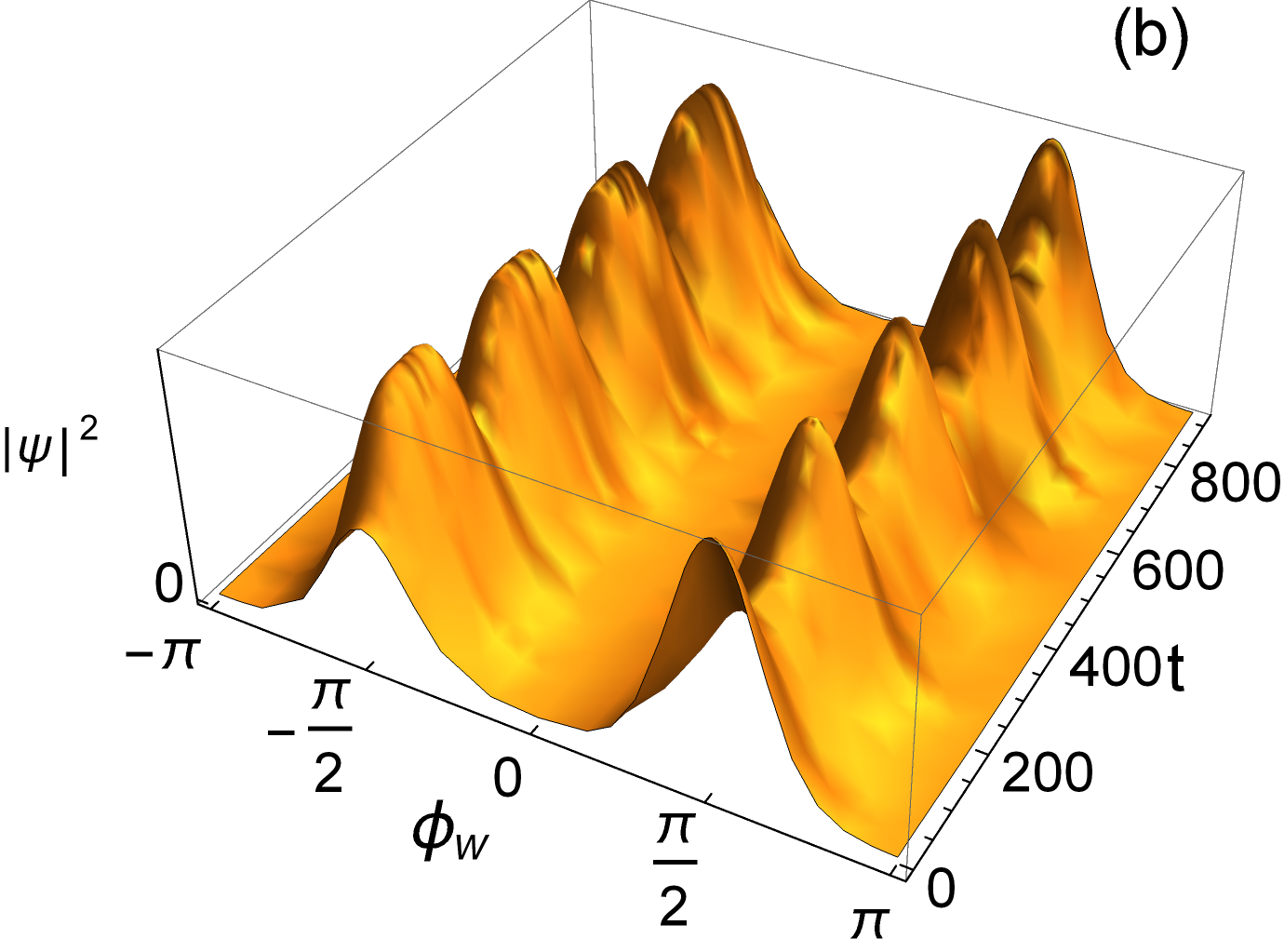}

\includegraphics[width=0.45\textwidth]{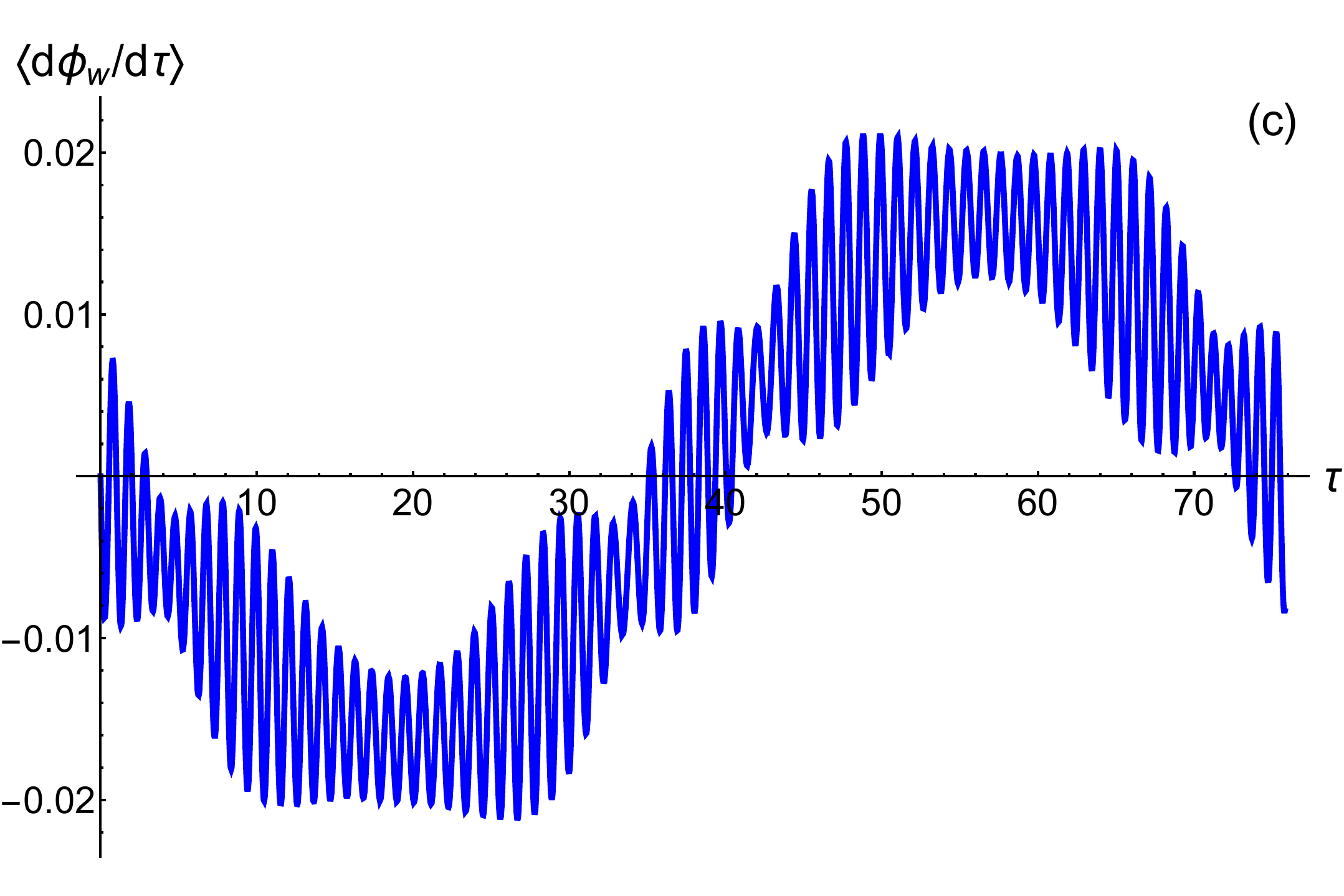}
\hspace*{5mm}\includegraphics[width=0.31\textwidth]{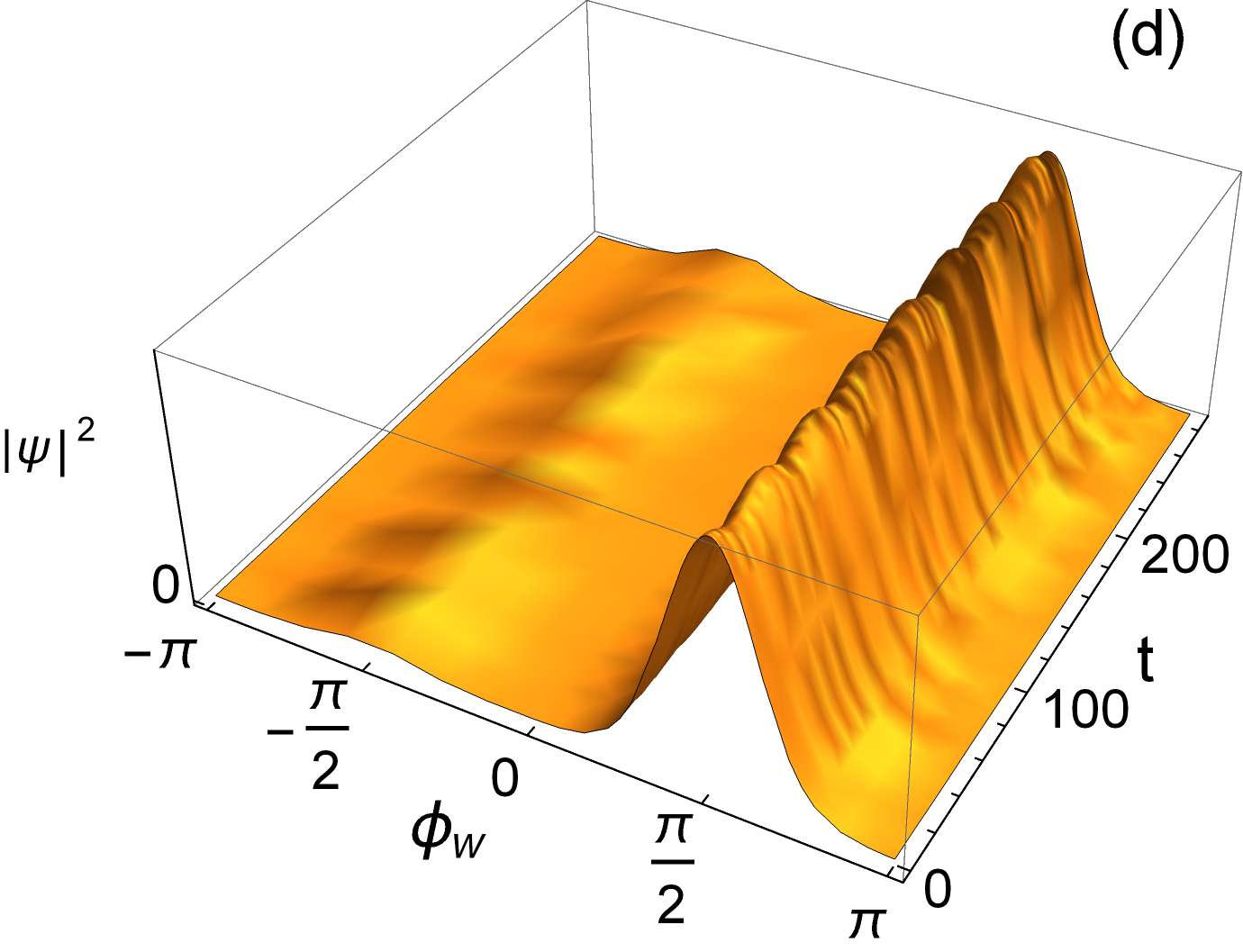}

\caption{\label{fig:bloch2L} 
(Color online).
Chirality dynamics with the Wannier state localized in the right well used as the initial state at $t=0$. Shown are: (a), (c) the average angular velocity $\langle d\phi_w/d\tau\rangle$, and (b), (d) the probability density $|\psi|^2$ as functions  of the dimensionless time $\tau=\hbar t/2m$  for $\xi=5$. The bias is  $b=0.1\Delta$ in (a), (b), and $b=0.6\Delta$ in  (c), (d) . 
 } 
\end{figure*} 
%%%%%%%%%%%%%%%%%%%%%%%%%%%%%%%%%%%%

\subsection{Classical chirality dynamics}
\label{subsec:class-switch}

The physics of this regime is essentially the same for FM, AF, and AM, and has been well studied for FM \cite{Martinez+2011} and AF \cite{Ovcharov+22}.  
Classically, the DW is ``stuck'' in one of the wells, and would be driven out only by a sufficiently large torque $b>b_c$, when the energy gain on traversing between wells becomes comparable to the barrier height, $b_c\pi\simeq U$. 
The threshold current density $j^{(c)}=I^{(c)}/C_\perp$, neccessary for such a switching, is given by
\begin{equation}
\label{jc-class}
j^{cl}=\frac{e E_0 }{2\pi\hbar a^2}\rho 
\end{equation}
and does not depend on the number of spins $N_\perp$ in the cross-section. 
As $U$ scales linearly with $N_\perp$,  one can simultaneously achieve low threshold current density 
 and high barrier $U\gg T$ (which is necessary to make the quasi-classical ``left'' and  ``right'' states stable against thermal fluctuations). 

In the absence of damping, a constant current above the threshold leads to a precession of the DW angle, with a constant acceleration. 
A finite damping  leads to a rotation with a stationary average angular velocity $\bar{\omega}\simeq  b/\hbar\eta$, where the dimensionless friction coefficient $\eta$ can be expressed via the Gilbert damping $\alpha_G$ as
\begin{equation}
\label{eta}
\eta=2N_\perp S \alpha_G (\delta/a).
\end{equation}
Such a setup has been suggested for nano-oscillator applications  \cite{OnoNakatani2008,Bisig+2009,Martinez+2011,ChengXiaoBrataas2016,Ovcharov+22}. 
Switching between two chiralities in this way, however, is plagued by inertial effects \cite{Ovcharov+22}: applying a pulse of current, one can kick the DW from one potential well, but it would not immediately stop in the other well, continuing the inertial rotation  until damping makes it settle.

\subsection{Quantum chirality dynamics}
\label{subsec:q-switch}

While classically, to excite the DW oscillations, one needs to apply the current density above the threshold value (\ref{jc-class}), 
quantum effects make it possible to excite the chirality dynamics  by much lower currents (i.e., torques much smaller than  the barrier height, $b\ll  U$). The key observation here is that the physics of a torque-driven quantum rotator is mathematically equivalent to that of a current-biased Josephson junction, which, in turn, is related to the Wannier-Stark problem for a charged particle in a periodic potential with an applied electric field (see the comprehensive review in \cite{Sonin22}). 
Drawing on this analogy, we show below that in the quantum regime the DW chirality dynamics can exhibit such effects as Bloch oscillations (with additional high-frequency modulation), Wannier-Stark localization, and Zener breakdown.

\paragraph{The Hamiltonian and the periodic gauge.}

The quantized form of the Hamiltonian (\ref{hamgen}) is easily obtained by promoting the conjugate momentum to operator 
$\hat{p}=-i\hbar\partial/\partial\phi_w$. 
It is convenient to introduce the dimensionless parameter 
\begin{equation}
\label{xi}
\xi= {2m\omega}/{\hbar}
\end{equation}
which determines the structure of the spectrum in the absence of the torque ($b=0$). 
For $\xi\gg1$, one has ``deep wells'', with the  low-energy spectrum being essentially that of a tunnel-split double harmonic oscillator with the level spacing $\hbar\omega$.  
The lowest-level tunnel splitting $\Delta$ and the harmonic oscillator level spacing $\hbar\omega$ can be related to the barrier height $U$ via the parameter $\xi$ as follows \cite{Sasaki82}:
\begin{equation}
\label{scales}
 \frac{\hbar\omega}{U}=\frac{4}{\xi},\quad \frac{\Delta}{U}= f(\xi),\quad f(\xi)=16\left(\frac{2}{\pi\xi}\right)^{1/2}e^{-\xi}.
\end{equation}
The above formula for $f(\xi)$ is derived for  for $\xi\gg1$ but remains a good approximation for $\xi \gtrsim 3$.
For energies considerably above $U$, the spectrum approximately corresponds to that of a free plane rotator. 
Decreasing $\xi$ pushes levels up, so that, e.g., for $\xi\sim 2.5\div 4$  only two lowest levels correspond to bound states inside the double well, and for $\xi\ll 1$ the entire spectrum is comprised by rotator-like delocalized states. 

%%%%%%%%%%%%%%%%%%%%%%%%%%%%%%%%%%%% 
\begin{figure}[tb] 
\includegraphics[width=0.45\textwidth]{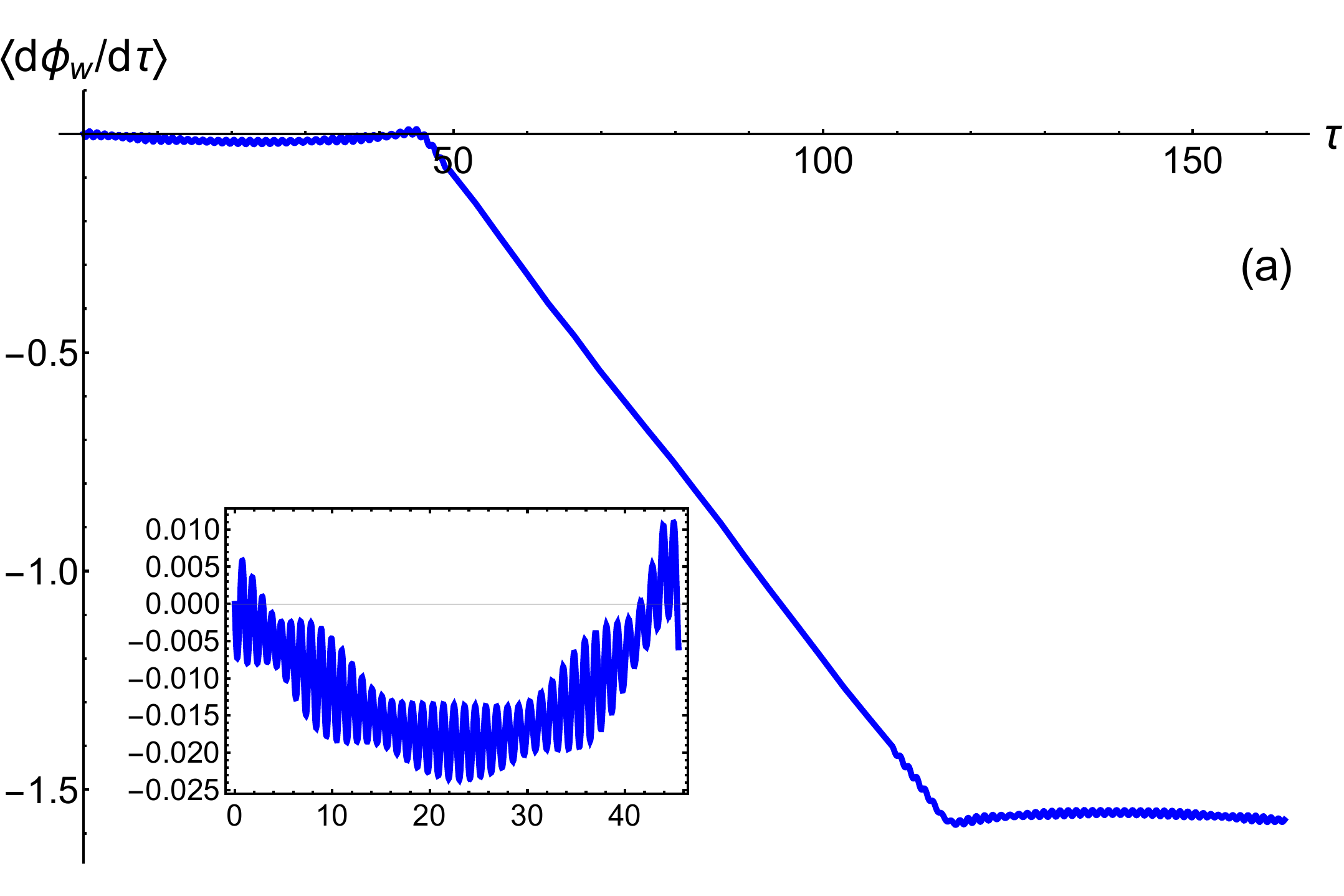}

\includegraphics[width=0.45\textwidth]{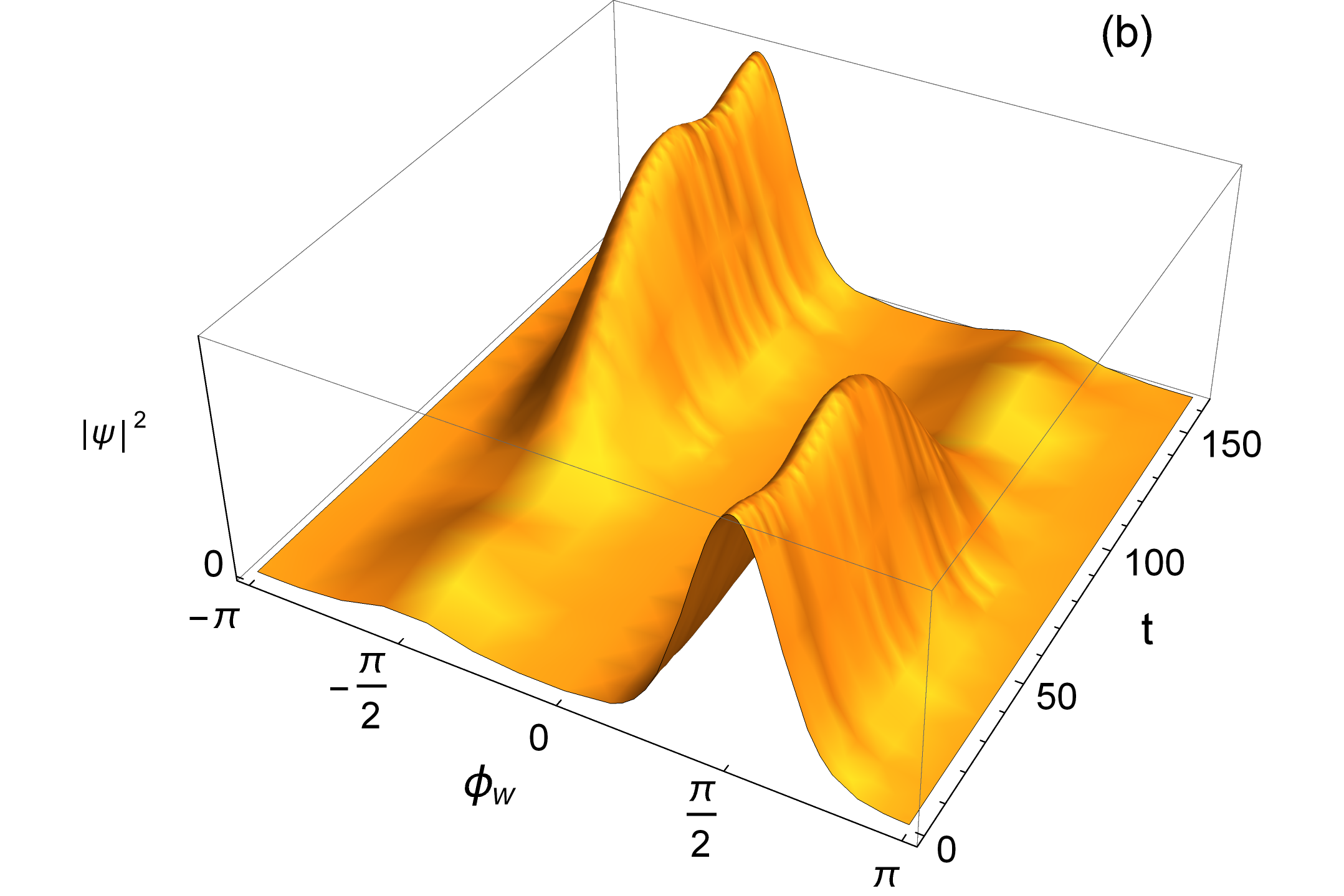}

\caption{\label{fig:switch} 
 (Color online).
Chirality switching with the Wannier state localized in the right well used as the initial state at $t=0$, under the action of a pulse with the amplitude $b=0.5\Delta$ and the duration of $\Delta t_1 =2\pi/\omega_B$, followed by another pulse of the same amplitude after the time interval of  $\Delta t_2 =\pi\hbar/\Delta$. 
Shown are: (a) the average angular velocity $\langle d\phi_w/d\tau\rangle$ and (b) the probability density $|\psi|^2$ as functions  of the dimensionless time $\tau=\hbar t/2m$  for $\xi=5$.
 } 
\end{figure} 
%%%%%%%%%%%%%%%%%%%%%%%%%%%%%%%%%%%%

To analyze the quantum chirality dynamics, it is convenient to pass from the wave function $\Psi(\phi_w)$ to a unitary-transformed one $\psi=e^{ik(t)\phi_w}\Psi$, with
\begin{equation}
\hbar k(t)=\int_0^t b(t')dt',
\end{equation}
which brings the Hamiltonian to the explicitly periodic, time-dependent form (the so-called ``periodic gauge'')
\begin{equation}
\label{ham-periodic}
\tilde\mathcal{H}= \frac{\hbar^2}{2m}\left\{\left( -i\frac{\partial}{\partial \phi_w} -k(t) \right)^2 +\frac{\xi^2}{4}\cos^2\phi_w \right\}.
\end{equation}
In this gauge one solves the time-dependent Schr\"odinger equation $i\hbar\partial_t \psi= \tilde\mathcal{H}\psi$ with periodic boundary conditions $\psi(\phi)=\psi(\phi+2\pi)$.

\paragraph{Bloch oscillations with high-frequency modulation.}

Consider first the case of a constant applied current, $\hbar k=bt$.
If the  energy gain on traversing between the wells is sufficiently small compared to the interband splitting, $ \pi b  \ll \hbar\omega$, the single-band approximation can be used \cite{LikharevZorin1985}. 
The angular velocity $\langle d\phi_w/dt\rangle$ should exhibit Bloch oscillations around zero average value, with the frequency
\begin{equation}
\label{omega-bloch}
\omega_B=\pi b/\hbar
\end{equation}
and the amplitude  $\bar\omega_m\simeq\pi\Delta/2\hbar$. The amplitude of the angle oscillations is thus $\phi_{\text{max}}\simeq \Delta/b$ and can be sufficiently large. While for short times $t\ll 1/\omega_B$ the Bloch oscillation is indistinguishable from the accelerated rotation, at  larger time scale they are very different as the average angular velocity  for the Bloch oscillation is zero.

Numerical solution of the time-dependent Schr\"odinger equation indeed shows such Bloch oscillations. 
Remarkably, we also observe a superimposed modulation with the frequency of about $\omega$, see Fig.\ \ref{fig:bloch1s}.  This can be explained by the fact that the matrix element $\langle \psi_3 |\hat{p}|\psi_0\rangle$ between the ground state $\psi_0$ and the third excited state $\psi_3$ (states at $b=0$)  is large, so nonzero $b$ leads to an efficient excitation of $\psi_3$ mode. 
 The amplitude of this additional modulation grows with increasing the current and gets more pronounced when the energy gain becomes comparable with the lowest level tunnel splitting.

In the presence of  a finite damping $\eta$, if the bias $b$ is lower than the threshold value $b_B=\hbar\eta \bar\omega_m\sim \eta\Delta$, there is a steady rotation with the angular velocity $\bar\omega=b/\hbar\eta$, similar to the classical regime.
The angular velocity starts Bloch oscillation for $b>b_B$, and its average (steady) value drops as $1/b$ \cite{Sonin22}.

\paragraph{The Wannier-Stark localization and chirality switching.}

In the regime of the energy gain not small compared to the bandwidth,  $ \pi b\gtrsim \Delta $, if the initial state is localized inside one of the wells, the Wannier-Stark localization will keep the wave function of the DW  localized, preventing it from tunneling, 
see Fig.\ \ref{fig:bloch2L}(d). Applying two current pulses separated by one half of the Rabi period $\pi\hbar/\Delta$, one can flip the localized state from one well to the other, see Fig.\ \ref{fig:switch}. Note that such a scheme for switching does not suffer from inertial effects that occur when trying to kick the DW from one well to the other in the classical regime. 
Thus, one can manipulate the chirality degree of freedom by the weak torque  $ \pi b \sim \Delta$, which corresponds to the operating current density  
\begin{equation}
\label{jc-quant}
j^{q}=j^{cl} f(\xi) ,
\end{equation}
which is considerably smaller than the classical switching threshold (\ref{jc-class})  for $\xi\gg 1$. 
The speed of quantum chirality switching is obviously limited by the tunnel splitting, with the maximum ``operating frequency'' about $\Delta/\hbar$.

\paragraph{Zener breakdown.}

\begin{figure*}[tb] 
\includegraphics[width=0.45\textwidth]{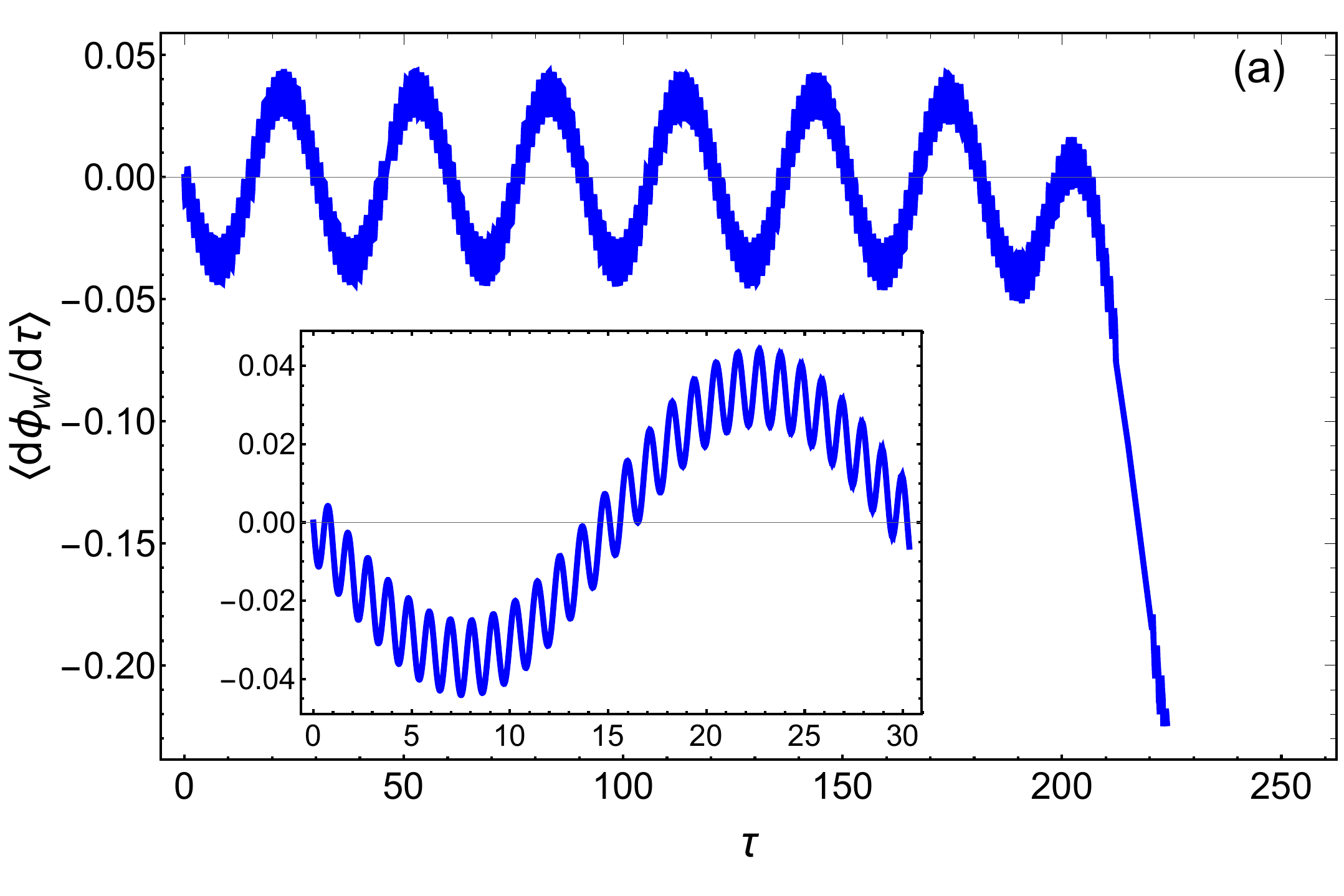}\hspace*{5mm} 
\includegraphics[width=0.32\textwidth]{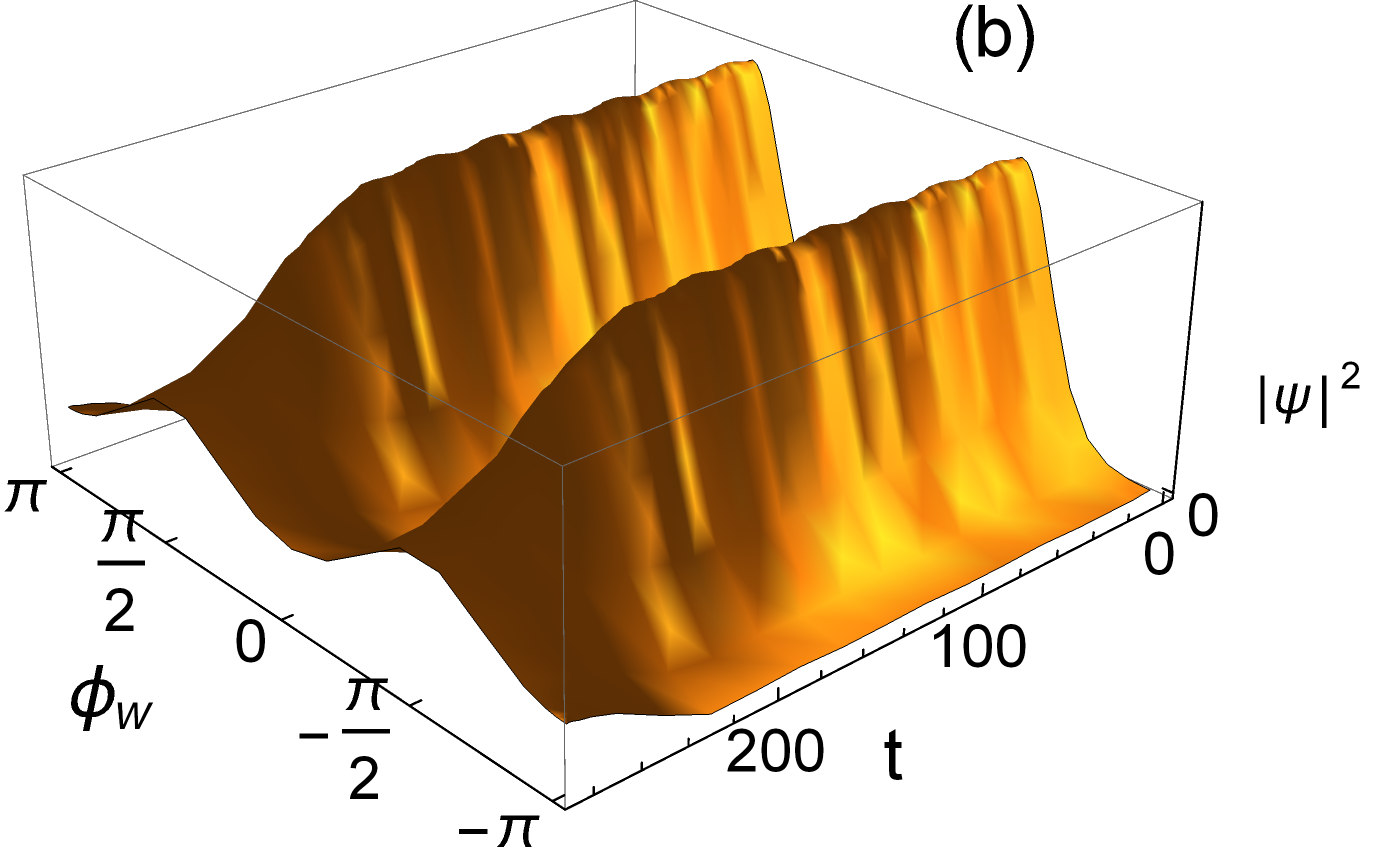}

\includegraphics[width=0.45\textwidth]{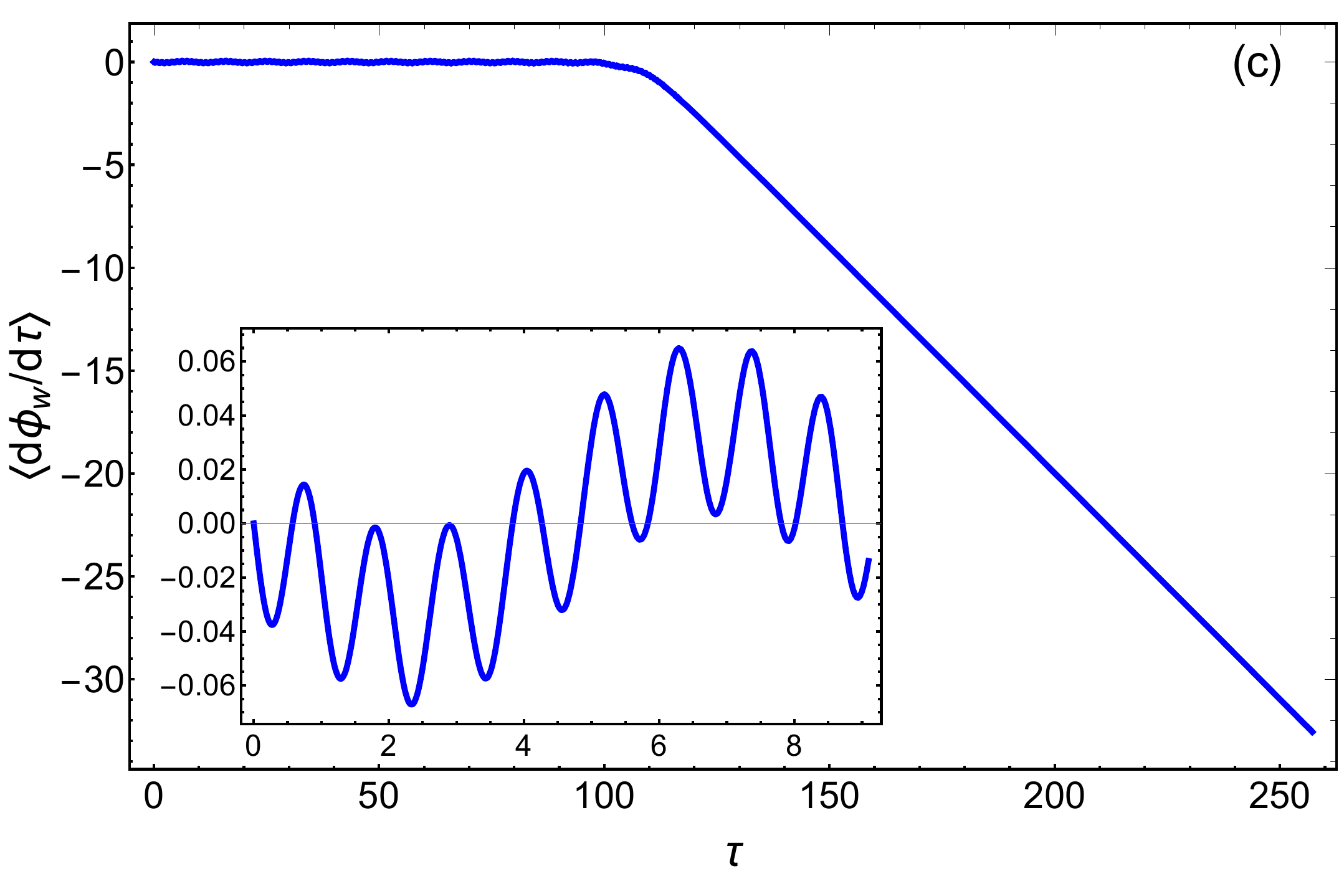}\hspace*{5mm} 
\includegraphics[width=0.32\textwidth]{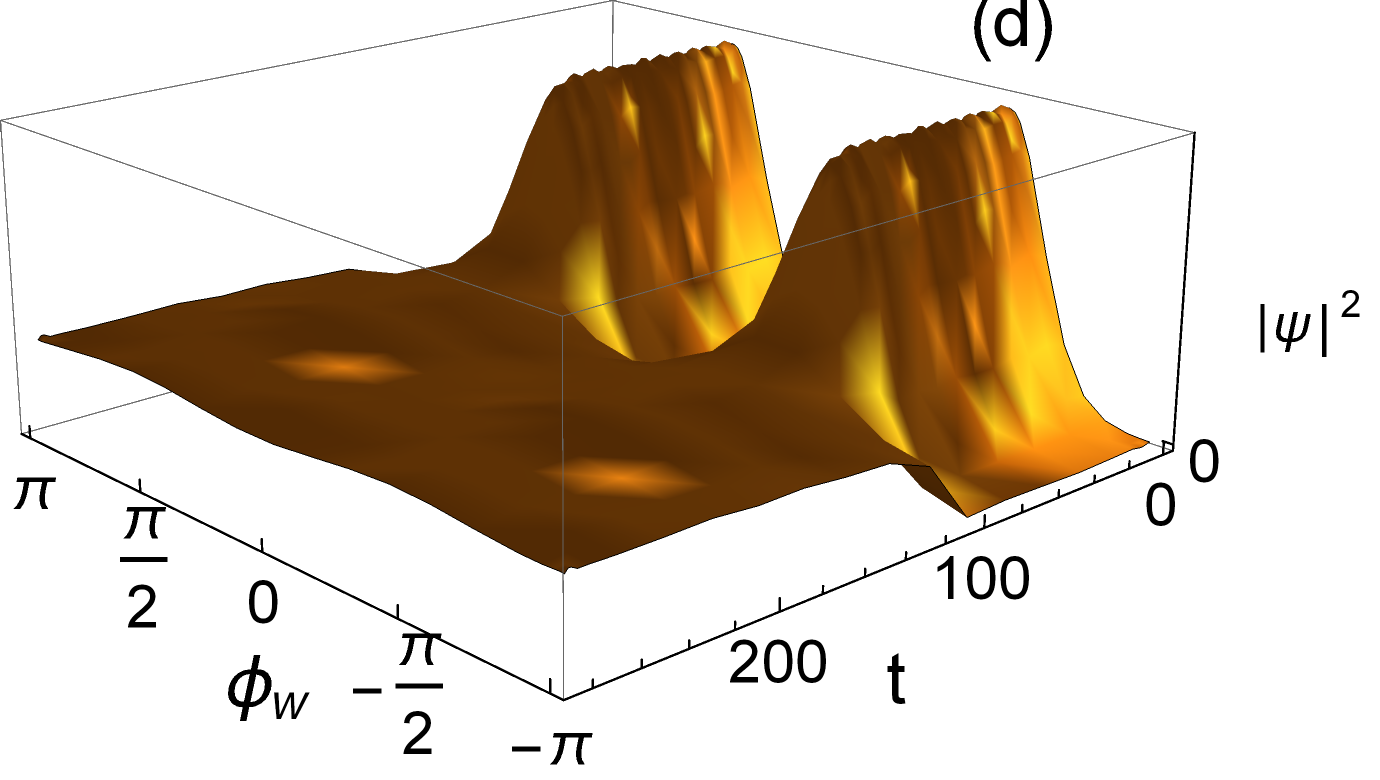}

\caption{\label{fig:zener} 
 (Color online).
Zener breakdown for a strong bias: (a), (c) the average angular velocity $\langle d\phi_w/d\tau\rangle$, and (b), (d) the probability density $|\psi|^2$ as functions  of the dimensionless time $\tau=\hbar t/2m$  for $\xi=5$. The bias is  $b=1.5\Delta$ in (a), (b), and $b=5\Delta$ in  (c), (d) .   The ground state for $b=0$ has been used as the initial state at $t=0$.
 } 
\end{figure*}

Finally, for strong bias the chirality dynamics can exhibit the Zener breakdown (interband tunneling) leading to an accelerated rotation, see Fig.~\ref{fig:zener}. The probability of the Zener breakdown becomes sizable when the energy gain becomes comparable  to the interband splitting, $\pi b\sim\hbar\omega$.  A finite damping would again change the dynamics to a steady rotation, so this regime is hardly distinguishable from the classical one. However,  for $\xi\gg1$ the typical current density $j^{\text{Zen}}\sim j^{cl} (4/\xi)$  required to trigger the breakdown   can be much lower than the classical threshold $j^{cl}$.

%%%%%%%%%%%%%%%%%%%%%%%%%%%%%%%%%%%% 
%%%%%%%%%%%%%%%%%%%%%%%%%%%%%%%%%%%%

\subsection{Numerical estimates}

To implement the quantum regime of chirality dynamics, several conditions should be met. Quantum tunneling starts to compete with thermal escape below the so-called ``crossover'' temperature $T_c$ which can be roughly estimated by equating the quantum tunneling rate $\Delta$ and the thermal escape rate $\hbar\omega e^{-U/T}$, which  yields
$T_c\sim U/\ln\left[ 4/\left( \xi f(\xi)\right)\right]$. 
To observe tunneling effects and Bloch oscillations, however, one needs to satisfy a stronger requirement \cite{LikharevZorin1985} of temperature being small compared to the bandwidth (the lowest-levels tunnel splitting), $T\ll \Delta\ll T_c$. Besides that, the Bloch oscillation regime requires the bias to be over the damping-related threshold $b>\eta\Delta$.  

The quantum reduction factor of the current necessary to manipulate the DW chirality is given by $j^{q}/ j^{cl}=\Delta/U=f(\xi)$.
Thus, to achieve the quantum reduction of the current necessary to manipulate the DW chirality, at the same time keeping the switching speed reasonably high, one has to work in the regime of moderate values of $\xi\sim 3-6$, 
so the number of magnetic atoms in the cross-section of the DW cannot be much larger than $N_\perp\sim 100$. To keep $\xi$ moderate, the rhombicity has to be small (nearly uniaxial anisotropy), but not too small as it would drive down the crossover temperature.  
To increase $T_c$ and $\Delta$, it is favorable  to have large exchange and anisotropy constants $J$, $K$ (although that simultaneously drives up $j^q$). 

Let us look at corresponding  order-of-magnitude numerical estimates,  assuming a nearly-uniaxial material with $\rho\sim 10^{-3}$, and taking typical values for the exchange constant $J\sim 500\,\text{K}$, the anisotropy constant $K\sim 5\,\text{K}$, spin $S=1$, and the lattice constant $a\sim 3$~\AA. The classical  current density threshold for the above parameters  is $j^{cl}\sim 10^5$~A/cm$^2$, independent of the number of atoms $N_\perp$ in the DW cross-section.

\paragraph{Antiferromagnets and altermagnets.} 

For an AF or AM,  parameter $\xi$ takes the form
\begin{equation}
\label{xi-af}
\xi=2N_\perp S\sqrt{\rho}
\end{equation}
which depends only on the rhombicity parameter $\rho$ and not on the overall magnitude of material constants (exchange coupling $J$ and magnetic anisotropy $K$). 

For a DW in a nanowire with  $N_\perp\sim 100$, one has $\xi\approx 6.3$ and the quantum reduction will bring the operation current down by two orders of magnitude, $j^{q}/ j^{cl}\sim 10^{-2}$. While the crossover temperature is $T_c\sim2\,\text{K}$, the bandwidth $\Delta\sim 100\,\text{mK}$ is rather low, limiting the experimental observation. 
Assuming the Gilbert damping $\alpha_G\sim10^{-3}$, the friction coefficient can be estimated as $\eta\sim 1$, so the threshold for exciting Bloch oscillations lies at about $j\sim j^{q}$. Bloch oscillations excited by $j \sim j^{q}$   have the frequency of $\sim 1\, \text{GHz}$, the maximum switching frequency being roughly of the same value. Those Bloch oscillations will be modulated by the much higher frequency of about $\omega_0\sim 0.1\,\text{THz}$.  As $\Delta$ is highly sensitive to the value of $\xi$, a slight reduction of the DW cross-section to $N_\perp\sim 64$ allows one to raise the bar on the observation temperature and the operating frequency, while still retaining a substantial quantum reduction: in that case, $\xi\approx 4$, which corresponds to $\Delta\sim 500\,\text{mK}$, operating frequency $\sim 10\, \text{GHz}$ and the reduction factor $j^{q}/ j^{cl}\sim 0.1$.

\paragraph{Ferromagnets.} 

For a domain wall in a FM,   
$\xi=4N_\perp^{3/2}(S/a)\left({\rho E_0}/{G}\right)^{1/2}$
apparently scales as $N_\perp^{3/2}$, however in fact the scaling is linear if one takes into account that the pinning constant $G$ must also scale linearly with $N_\perp$. Following  \cite{TakagiTatara96}, one can model the pinning as caused by $N_p$ layers of impurities with an increased anisotropy constant $K'>K$ and obtain the estimate $G\sim N_p N_\perp K'/\delta^2$, which yields
$\xi=4N_\perp S^2 \left(\frac{\rho J}{N_p K'}\right)^{1/2}\left(\frac{2 J}{K}\right)^{1/4}$.
Thus, in FM $\xi$ is  enhanced by the large factor $({\delta}/{a})\left({E_0}/{N_p K'}\right)^{1/2}$
 compared to the AF case (\ref{xi-af}), which makes it unfeasible to observe Bloch oscillations and other quantum effects in a FM.

%%%%%%%%%%%%%%%%%%%%%%%%%%%%%%%%%%
\section{Summary} 
\label{sec:summary} 
%%%%%%%%%%%%%%%%%%%%%%%%%%%%%%%%%%

We have studied different scenarios of manipulating the internal degree of freedom (chirality) of a pinned mesoscopic domain wall by means of applying external current, in several types of magnetic materials.   
While one needs spin-polarized current to control chirality in ferro- and antiferromagnets, it turns out that in the emerging novel class of magnetic materials known as altermagnets chirality can be manipulated by a regular, non-polarized charge current.

We have shown that in antiferromagnets or altermagnets 
at very low temperatures, quantum tunneling effects open the possibility for  controlling the chirality dynamics by 
current densities which can be an order of magnitude smaller than those required to excite it classically. Those very weak currents can excite quantum Bloch oscillations of the domain wall angular velocity with frequencies of the order of a few GHz, modulated by oscillations at much higher frequencies in the THz domain.  Zener breakdown can be triggered by higher currents that are still substantially lower than the classical threshold. Beside exciting the oscillations, in the quantum regime it is possible to perform a controlled switching between different chiral states, with inertial effects suppressed by the Wannier-Stark localization, albeit with the payoff of a limited switching speed.

\begin{acknowledgments}
We thank J. Sinova, M. Kl\"aui, L. \v{S}mejkal, and Y. Bazaliy for fruitful discussions.  
O.G. acknowledges support by the Deutsche Forschungsgemeinschaft (DFG, German Research Foundation) via TRR 173–268565370 (project A11) and TRR 288–422213477. 
O.K. gratefully acknowledges support by the Philipp Schwartz Initiative, and the hospitality of the Johannes Gutenberg University of Mainz. 
\end{acknowledgments}

%%%%%%%%%%%%%%%%%%%%%%%%%%%%% %

\appendix* 
\section{Current coupling to the order parameter in ferro-, antiferro- and altermagnets} 
\label{app} 

%%%%%%%%%%%%%

The goal of this Appendix is to give a derivation of the current coupling to the order parameter in altermagnets \cite{Smejkal+22rev}, but we would like to provide the necessary context by giving a brief overview of ferro- and antiferromagnetic cases  first. 

It is convenient to start from a tight-binding model for conduction electrons described by the Hamiltonian
\begin{equation}
\label{tight}
H_{sd}=-t\sum_{\langle ll' \rangle}(\psi_l^\dagger \psi_{l'}^{\vphantom{\dagger}} +\text{h.c.}) -J_{sd}\sum_l \psi_l^\dagger (\vec{\sigma}\cdot\vec{S}_l) \psi_l^{\vphantom{\dagger}},
\end{equation}
where $l$ numbers lattice sites, $\psi_l=(c^\uparrow_l,c^\downarrow_l)^T$ is the two-component spinor  describing the conduction electrons (in the frame with the quantization axis $\vec{n}_0$), $t$ is the hopping amplitude between nearest neighbor site pairs $\langle ll' \rangle$, and $J_{sd}$ is the exchange coupling to localized spins $\vec{S_l}$ which are treated as classical vectors. The Lagrangian corresponding to this tight-binding model can be written as
\begin{equation}
\label{tight-L}
L_{sd}=i(\hbar/2)\sum_l \left( \psi_l^\dagger \partial_t \psi_{l}^{\vphantom{\dagger}} - \partial_t \psi_l^\dagger \psi_{l}^{\vphantom{\dagger}}\right) -H_{sd} .
\end{equation}

\paragraph*{Ferromagnet.}

In a ferromagnet, $\vec{S_l}=S\vec{n}_l$, where the unit vector $\vec{n}$ can be assumed to vary smoothly across the lattice. Performing local unitary transformation $\psi_l =U(\vec{n})\chi_l$ with 
\begin{equation}
\label{unitary}
U(\vec{n}_l)=\vec{\sigma}_l\cdot\vec{e}_l,\quad 
\vec{e}=\frac{ \vec{n}_0+\vec{n}}{[2(1+\vec{n}\cdot\vec{n}_0)]^{1/2}},
\end{equation}
which describes a $180^{\circ}$ rotation about the direction $\vec{e}$ that bisects the angle between $\vec{n}_l$ and $\vec{n}_0$,
we rotate the quantization axis at each lattice site to $\vec{n}_l$, so that the interaction term $\psi_l^\dagger (\vec{\sigma}\cdot\vec{S}_l) \psi_l =S \chi_l^\dagger \sigma_z \chi_l$ is diagonalized in this twisted frame. This twist modifies hopping as follows:
\begin{equation}
\label{hopping}
\psi_l^\dagger \psi_{l'}  = (\vec{e}_l\cdot\vec{e}_{l'}) \chi_l^\dagger \chi_{l'} +  i (\vec{e}_l\cdot\vec{e}_{l'})\cdot (\chi_l^\dagger \vec{\sigma} \chi_{l'}). 
\end{equation}
One can pass to the continuum description, by setting $l\mapsto \vec{r}$, $l'\mapsto\vec{r}+\vec{a}$, and perform the gradient expansion. Up to the second order in gradients, the hopping term in (\ref{tight}) takes the form
\begin{equation}
\label{kinetic}
ta^2(\nabla_a \psi^\dagger)(\nabla_a\psi) = ta^2 \big(\nabla\chi^\dagger -i\chi^\dagger \mathsf{A}_a\big)
\big(\nabla\chi +i \mathsf{A}_a\chi\big),
\end{equation}
where the matrix gauge field $\mathsf{A}_a=(\vec{\mathcal{A}}_a\cdot\vec{\sigma})$ is defined via the set of vector gauge fields
\begin{equation}
\label{gauge-vec}
 \vec{\mathcal{A}}_a = (\vec{e}\times \nabla_a\vec{e})=\frac{(\vec{n}_0+\vec{n})\times\nabla_a \vec{n}}{2(1+\vec{n}\cdot\vec{n}_0)},
\end{equation}
and we have used the shorthand notation $\nabla_a \equiv \frac{1}{a}(\vec{a}\cdot\vec{\nabla})$. 

The  expression (\ref{kinetic}) describes the kinetic energy of electrons with the effective mass 
\begin{equation}
\label{eff-mass}
m_*=\hbar^2/(2ta^2)
\end{equation}
coupled to SU(2) gauge field $\mathsf{A}$. In Eq.\ (\ref{kinetic}), the term quadratic in $\mathsf{A}$ can be recast as $ta^2 (\chi^\dagger \chi) (\nabla \vec{e})^2$ and leads to a renormalization of the FM exchange energy in the presence of a finite density of conduction electrons, so we will omit it in what follows. The term linear in $\mathsf{A}$ yields the following contribution to the Lagrangian density in the continuum approximation:
\begin{equation}
\label{coupling-spincurrent}
-2\vec{j}^{(s)}_a\cdot\vec{\mathcal{A}}_a,
\end{equation}
where 
\begin{equation}
\label{spin-current}
\vec{j}^{(s)}_a =-\frac{i\hbar^2}{4m_*V_0}\big( \chi^\dagger\vec{\sigma}\nabla_a\chi -\nabla_a \chi^\dagger\vec{\sigma}\chi \big)
\end{equation}
is the spin current density in the $\vec{a}$ direction  (the components of the vector $\vec{j}^{(s)}_a$ are taken in the rotated frame), and $V_0$ is the volume of the magnetic unit cell. 

The first term in the Lagrangian (\ref{tight-L}), after performing the unitary twist, similarly yields another contribution to the Lagrangian density 
\begin{equation}
\label{coupling-spinacc}
-2\vec{\tau}^{(s)}\cdot\vec{\mathcal{A}}_0,
\end{equation}
where the gauge field $\vec{\mathcal{A}}_0$ has the form (\ref{gauge-vec}) with the replacement $\nabla_a\mapsto\partial_t$, and  
$\vec{\tau}^{(s)}$ is the spin density of the conduction electrons in the locally rotated frame (i.e., with respect to the order parameter):
\begin{equation}
\vec{\tau}^{(s)}= (\hbar/2V_0)\left( \chi^\dagger\vec{\sigma}\chi \right).
\end{equation}
Integrating (\ref{coupling-spincurrent}) and (\ref{coupling-spinacc}) over the volume of the nanowire, one gets the total current-induced contribution to the Lagrangian of the form
\begin{equation}
\label{scurr-cpl}
\mathcal{L}\mapsto \mathcal{L}-2C_\perp\int dz\,\left( \vec{j}^{(s)}\cdot \vec{\mathcal{A}}_z + \vec{\tau}^{(s)}\cdot \vec{\mathcal{A}}_t \right) .
\end{equation}
We assume that the exchange coupling between conduction electrons and localized spins is strong enough so the electron spin adiabatically follows the direction of the magnetization (so the coupling (\ref{scurr-cpl}) corresponds to the adiabatic spin-transfer torque). Then in the locally rotated frame $\vec{j}^{(s)}=j^{(s)} \hat{\vec{z}}$ and $\vec{\tau}^{(s)}=\tau^{(s)} \hat{\vec{z}}$, so 
one can recast (\ref{scurr-cpl}) in the form of Eq.\ (\ref{scurr-cpl1}), 
where $I^{(s)}=j^{(s)} C_\perp$, $\mathcal{T}^{(s)}=\tau^{(s)} C_\perp$.

 While $P_z=\int \vec{A}\cdot d\vec{n}$ itself is not a well-defined quanity,  the difference of $P_z$ values for two domain walls  with $\varphi=\phi_1$ and $\varphi=\phi_2$ is uniquely defined \cite{Galkina+08chir} since it can be expressed as the surface integral $\int dS\, (\vec{n}\cdot\text{rot}_{\vec{n}} \vec{A}) $ over the spherical wedge of the $\vec{n}$ unit sphere,  bounded by the semidisks   $\varphi=\phi_{1,2}$. 
 In view of the identity $\vec{n}\cdot \text{rot}_{\vec{n}} \vec{A}=1$,  this latter quantity is just equal to the surface $2(\phi_1-\phi_2)$ of the wedge. Thus, up to an irrelevant constant, we can set $P_z=2\phi_w$, obtaining (\ref{Pw}).

\paragraph*{Antiferromagnet.}

In a two-sublattice antiferromagnet,  $\vec{S_l}=S(\vec{m}_l+\eta_l\vec{n}_l)$, where $\eta_l$ takes alternating values $\pm1$ on two sublattices, and $\vec{m}$ and $\vec{n}$ correspond to the magnetization and the N\'eel vector, respectively, satisfying the constraints $\vec{m}\cdot\vec{n}=0$ and $\vec{m}^2+\vec{n}^2=1$.  For smooth spin textures, the magnetization  $\vec{m}$ can be viewed as a ``slave field" which is proportional to  space and time gradients of the order parameter $\vec{n}$,  so typically  $|\vec{m}|\ll|\vec{n}|$, and one can for our purposes neglect $\vec{m}$ and view $\vec{n}$ as a unit vector. Performing the same unitary transformation (\ref{unitary}) adjusts the electron quantization axes to the local N\'eel vector direction and brings the interaction term to its form for the homogeneous AF texture. 
The hopping term produces \cite{SwavingDuine11} the  gauge-field coupling (\ref{coupling-spincurrent}), while 
 the dynamic term in the Lagrangian produces \cite{ChengNiu14}  the gauge-field couplings (\ref{coupling-spinacc}), similar to the FM case. 
We remark that our derivation here is kept simplified for the sake of clarity. The more rigorous derivation \cite{ChengNiu12,ChengNiu14}  leads to the additional overall factor $(1-\zeta^2)$ in front of both (\ref{coupling-spincurrent}) and (\ref{coupling-spinacc}), with $\zeta\propto t/J_{sd}$. Here we assume that $\zeta\ll 1$ and  neglect this additional factor.

\paragraph*{Altermagnet.}

As a starting point, we take one of the minimal toy models of an altermagnet \cite{Smejkal+22rev}, described by the Hamiltonian (\ref{tight-am}) and Fig.\ \ref{fig:sqlatam}. In each unit cell we introduce the magnetization $\vec{m}$ and the N\'eel vector $\vec{n}$: 
\begin{equation}
\label{ml-am}
\vec{S}_{1,l}=S(\vec{m}_{l}+\vec{n}_{l}),\quad \vec{S}_{2,l}=S(\vec{m}_{l}-\vec{n}_{l}).
\end{equation}
In what follows, we assume that in the homogeneous unperturbed state $\vec{m}=0$, and $\vec{l}=\vec{n}_0$, and $\vec{n}_0$ is also the quantization axis for electron spinors $\psi_l$. 

We again perform unitary transformation (\ref{unitary}) to rotate the electron quantization axes to the local N\'eel vector direction $\vec{n}_l$. The spin-independent term in the hopping will again acquire the gauge-field modification (\ref{kinetic}), leading to the contribution of the form  (\ref{coupling-spincurrent}) coupling texture gradients to the spin current. However, it is easy to show that the spin-dependent hopping leads to another contribution which couples not to the spin current, but to the charge current. 
Indeed, the $t'$-term transforms as follows: $\psi_l^\dagger (\vec{\sigma}\cdot\vec{S}_{ll'}) \psi_{l'} = \chi_l^\dagger W_{ll'} \chi_{l'}$, where
\begin{eqnarray}
\label{unitary-am}
W_{ll'}&=& -i \vec{S}_{ll'}\cdot (\vec{e}_l\times \vec{e}_{l'})\;\mathds{1} \\
&+&
\Big\{\vec{e}_{l}(\vec{e}_{l'}\cdot\vec{S}_{ll'}) + \vec{e}_{l'}(\vec{e}_{l}\cdot\vec{S}_{ll'}) - \vec{S}_{ll'}\cdot (\vec{e}_{l}\cdot\vec{e}_{l'})\Big\}\cdot\vec{\sigma}.\nonumber
\end{eqnarray}
Passing to the continuum description, $l\mapsto \vec{r}$, $l'\mapsto\vec{r}+\vec{a}$, and performing the gradient expansion, one can see that
up to the first order in gradients $W$ takes the following form:
\begin{eqnarray}
\label{hopping-am}
W&=&\eta_{\vec{a}} (\vec{n}_0\cdot\vec{\sigma}) + \Big( 2\vec{e}(\vec{e}\cdot \vec{m})-\vec{m}\Big)\cdot \vec{\sigma} \nonumber\\
&+&\eta_{\vec{a}} a  \Big( \vec{e}(\vec{n}\cdot\nabla_a\vec{e})+\nabla_a\vec{e} (\vec{n} \cdot\vec{e}) \Big) \cdot\vec{\sigma} \\
&-&i a\vec{n}\cdot \big( \vec{e}\times \nabla_a \vec{e}) \mathds{1},\nonumber
\end{eqnarray}
where $\eta_{\vec{a}}=\pm 1$ for $\vec{a}=\hat{\vec{x}}$, and $\vec{a}=\hat{\vec{y}}$, respectively.
Similar to the low-energy description of antiferromagnets, the magnetization  $\vec{m}$ is linear in gradients of the order parameter $\vec{n}$.
The gradient-free term in $W$ corresponds to the spin-dependent hopping in the homogeneous AF order. 

To the first order in gradients, there are two contributions to $W$. One of them is a superposition of Pauli's matrices, and describes coupling of the electron spin density to the gradient of the order parameter; in what follows, we will not be interested in that contribution as it does not involve currents. The other term in $W$, which is purely imaginary and proportional to the unit matrix, leads to the following contribution to the Lagrangian density in the continuum approximation:
\begin{equation}
\label{coupling-ccurrent}
-\frac{t'}{t}\frac{\hbar}{e}\sum_{\vec{a}=\hat{\vec{x}},\hat{\vec{y}} } j^{(c)}_{a} \eta_{\vec{a}} (\vec{n}\cdot\vec{\mathcal{A}}_{a}),
\end{equation}
where 
\begin{equation}
\label{ccurrent}
j^{(c)}_{a} =-\frac{ie\hbar}{2m_* V_0}\big( \chi^\dagger\nabla_a\chi -\nabla_a \chi^\dagger\chi \big)
\end{equation}
is the density of the charge current in the direction $\vec{a}$. Thus,  in an altermagnet the N\'eel vector gradient couples to the usual electric current, in contrast to the FM or AF case, where coupling is only to the spin-polarized current. Taking into account (\ref{gauge-vec}), the coupling (\ref{coupling-ccurrent}) can be recast as 
\begin{equation}
\label{alt-coupling}
-\kappa\frac{\hbar}{2e}\sum_{\vec{a}=\hat{\vec{x}},\hat{\vec{y}} } j^{(c)}_{a} \eta_{\vec{a}} \vec{A}(\vec{n}_0,n)\cdot\nabla_a \vec{n},
\end{equation}
where the Dirac monopole  vector potential $\vec{A}(\vec{n}_0,n)$ is defined in (\ref{Adirac}), and $\kappa=t'/t$ is the relative strength of the spin-dependent hopping; this leads to (\ref{am-coupling}).

Finally, the contribution from the dynamic part of the Lagrangian (\ref{tight-L}) will result in the same coupling term (\ref{coupling-spinacc}). 
If the current is not spin-polarized, this contribution vanishes, and the only surviving contribution is (\ref{alt-coupling}).

%%%%%%%%%%%%%%%%%%%%%%%%%%%%%%%%%%%%%%%%%%%%%%%%%%%%%%%%%%%% 

\bibliography{dw-chir}

%%%%%%%%%%%%%%%%%%%%%%%%%%%%%%%%%%%%%%%%%%%%%%%%%%%%%%%%%%%%%

\end{document}